\documentclass[aip,reprint]{revtex4-1}

\usepackage{epsfig}
\usepackage{color}
\usepackage{amsmath}
\usepackage{amssymb}
\usepackage{longtable}
\usepackage{dcolumn}
\usepackage{bm}
\usepackage{ulem}
\usepackage{float}
\usepackage{mathtools}
\usepackage{xparse}
\NewDocumentCommand{\ceil}{s O{} m}{%
  \IfBooleanTF{#1} 
    {\left\lceil#3\right\rceil} 
    {#2\lceil#3#2\rceil} 
}

\newcommand{\ket}[1]{\mathinner{|{#1}\rangle}}

\newcommand{\braket}[2]{\langle #1|#2\rangle}

\providecommand{\abs}[1]{\lvert#1\rvert}

\draft 

\begin{document}


\title{A Quantum Annealing Approach for Fault Detection and Diagnosis of Graph-Based Systems} 



\author{Alejandro Perdomo-Ortiz}
\email[Corresponding author's e-mail: ]{$\,$alejandro.perdomoortiz@nasa.gov}
\affiliation{Quantum Artificial Intelligence Lab.,
NASA Ames Research Center,
Moffett Field, CA 94035, USA}
\affiliation{University of California Santa Cruz @
NASA Ames Research Center, Moffett Field, CA 94035, USA}


\author{Joseph Fluegemann}
\affiliation{Quantum Artificial Intelligence Lab.,
NASA Ames Research Center,
Moffett Field, CA 94035, USA}
\affiliation{San Jose State Research Foundation @ NASA Ames Research Center, Moffett Field, CA 94035, USA}


\author{Sriram Narasimhan}
\affiliation{University of California Santa Cruz @
NASA Ames Research Center, Moffett Field, CA 94035, USA}


\author{Rupak Biswas}

\affiliation{Quantum Artificial Intelligence Lab.,
NASA Ames Research Center,
Moffett Field, CA 94035, USA}


\author{Vadim N. Smelyanskiy}
\affiliation{Quantum Artificial Intelligence Lab.,
NASA Ames Research Center,
Moffett Field, CA 94035, USA}


\date{\today}


\maketitle 


\textbf{Diagnosing the minimal set of faults capable of explaining a set of given observations, e.g., from sensor readouts, is a hard combinatorial optimization problem usually tackled with artificial intelligence techniques. We present the mapping of this combinatorial problem to quadratic unconstrained binary optimization (QUBO), and the experimental results of instances embedded onto a quantum annealing device with 509 quantum bits. Besides being the first time a quantum approach has been proposed for problems in the advanced diagnostics community, to the best of our knowledge this work is also the first research utilizing the route Problem $\rightarrow$ QUBO $\rightarrow$ Direct embedding into quantum hardware, where we are able to implement and tackle problem instances with sizes that go beyond previously reported toy-model proof-of-principle quantum annealing implementations; this is a significant leap in the solution of problems via direct-embedding adiabatic quantum optimization. We discuss some of the programmability challenges in the current generation of the quantum device as well as a few possible ways to extend this work to more complex arbitrary network graphs.}


\section{Introduction}

Electrical power-distribution systems (EPS) are ubiquitous, and in many instances, their reliability is critical for the success of a mission. We focus here on the diagnosis of multiple faults~\cite{deKleer1987} in EPS where the problem is to determine which components are in a failed state, given observations from sensors placed, for example, on an aircraft. Typical model-based approaches use heuristic-driven search over the component failure space to try and determine the failed components. In general this is a hard problem, and the complexity grows exponentially with problem size and with the number of possible failed components. The goal of this work is to use the D-Wave quantum machine to perform the search and compare its solutions with classical state-of-the-art technologies for these specific problems. One such classical methodology developed at NASA is the Hybrid Diagnostics Engine (HyDE)~\cite{narasimhan2007hyde}, which we are using for performance comparison against our quantum algorithms.

Harnessing quantum-mechanical effects to speed up the solving of classical optimization problems is at the heart of quantum annealing algorithms (QA)~\cite{finnila_quantum_1994,kadowaki_quantum_1998,santoro_optimization_2006,das_2008,ray1989}. There is theoretical~\cite{amara_global_1993,finnila_quantum_1994,kadowaki_quantum_1998,Farhi2001,santoro_theory_2002}  and experimental~\cite{brooke_quantum_1999} evidence of scenarios where using QA~\cite{finnila_quantum_1994,kadowaki_quantum_1998,santoro_optimization_2006,das_2008} to solve classical optimization problems could be advantageous over its classical analogue (simulated annealing~\cite{kirkpatrick_optimization_1983}). For the case of the D-wave quantum processor, several benchmark studies~\cite{McGeoch2013,boixo_evidence_2014,Ronnow2014_quantumspeedup} have recently been reported with, as yet, inconclusive results about the quantum speedup that do not rule out the possibility of finding applications or problem instances where one could take advantage of a QA algorithm. The search for scenarios where this advantage can be harnessed is still an open question~\cite{katzgraberPRX2014,Venturelli2014,GoogleQuantumAI}. In QA, quantum mechanical tunneling allows for more efficient exploration of difficult potential energy landscapes such as that of classical spin-glass problems. In our implementation of the diagnosis of multiple faults, quantum fluctuations (tunneling) occurs between states representing different diagnosis candidates.

Despite the significant progress and exponential growth in the number of qubits in the commercial D-wave quantum annealer, implementing and solving practical applications in these devices is still a challenging task, as discussed in this work and other recently developed QA algorithms in other applications domains~\cite{perdomo08,PerdomoOrtiz2012_LPF,planningquantum,ogorman2014_bayesnet}. When compared to traditional computers or workstations, with billions of bits, the main bottleneck for quantum annealers is still the size of the device, which now is in the 512 quantum bit (qubit) prototype, or more precisely, 509 functional qubits in the quantum processor acquired through the NASA-Google-USRA partnership. Implementing a problem in the language of the quantum machine requires a mapping of the problem to a quadratic unconstrained binary optimization problem (QUBO), and the subsequent embedding into the hardware architecture (see Fig.~\ref{fig:diagnosis}), both steps usually demanding more bits than the bits needed to solve the problem with a classical algorithm.

These constraints have limited the implementations of practical problems to proof-of-principle demonstrations~\cite{Gaitan2012,PerdomoOrtiz2012_LPF,planningquantum}. In this paper we present the first application with the route Problem $\rightarrow$ QUBO $\rightarrow$ Direct embedding into quantum hardware, where we are able to implement and tackle problem instances with sizes that not only require at least a laptop to find the solution but also their number of elements comparable to those found in real-world problems. For example, in the electrical circuits used for diagnosis competitions from NASA's Advanced Diagnostics and Prognostics Testbed (ADAPT), this number ranges between 40-100 components~\cite{kurtoglu09first}. In the present paper, we can easily embed into the D-wave machine, problem instances with close to 100 components, split among circuit breakers and light-emitting diodes. We believe that these results represent a significant leap in the solution of problems via direct-embedding quantum optimization.


In Sec.~\ref{sec:quantumDMF} we describe all the necessary steps involved in implementing a hard computational problem into the quantum machine, starting from the mapping to a QUBO, and the subsequent embedding in the quantum hardware. In Sec.~\ref{sec:ResultsDiscussion} we present and discuss the results of our preliminary experimental runs on the quantum annealer. Finally, in Sec.~\ref{sec:conclusion} we provide some concluding remarks and an outlook on the application of quantum algorithms in system health management.

\section{Quantum annealing approach for the diagnosis of multiple faults problem}\label{sec:quantumDMF}

Our approach to the problem is to use the D-Wave quantum computer and exploit quantum fluctuations, i.e., quantum tunneling, to more efficiently explore the search space. The theoretical challenge is to efficiently map the hard computational problem of interest (e.g., diagnosis of multiple faults) to a quadratic pseudoboolean objective function (classical spin-glass Hamiltonian in physics jargon) to be minimized; such a mapping requires that the number of qubits scales polynomially with the size of the problem (number of electrical components which are potentially faulty). Solving arbitrary problem instances requires a programmable quantum device to implement the corresponding QUBO. We employ quantum annealing on the programmable device to diagnose faults in electrical power networks.

The QA protocol performed here is also known as adiabatic quantum computation (AQC)~\cite{farhi2000,Farhi2001}. Of all the quantum-computational models, AQC is perhaps the most naturally suited for studying and solving optimization problems~\cite{Farhi2001,hogg03}. 
In this section, we describe in detail the problem of interest, and provide the mapping of this problem to a QUBO expression. Once in this form, one can embed the final QUBO expression into the quantum hardware, which possesses a fixed topology, thus requiring an overhead in the resources, as explained in further detail in Sec.~\ref{sec:ResultsDiscussion}.

\subsection{Quantum Annealing}

\begin{figure}
\centering
\includegraphics[width=0.45\textwidth]{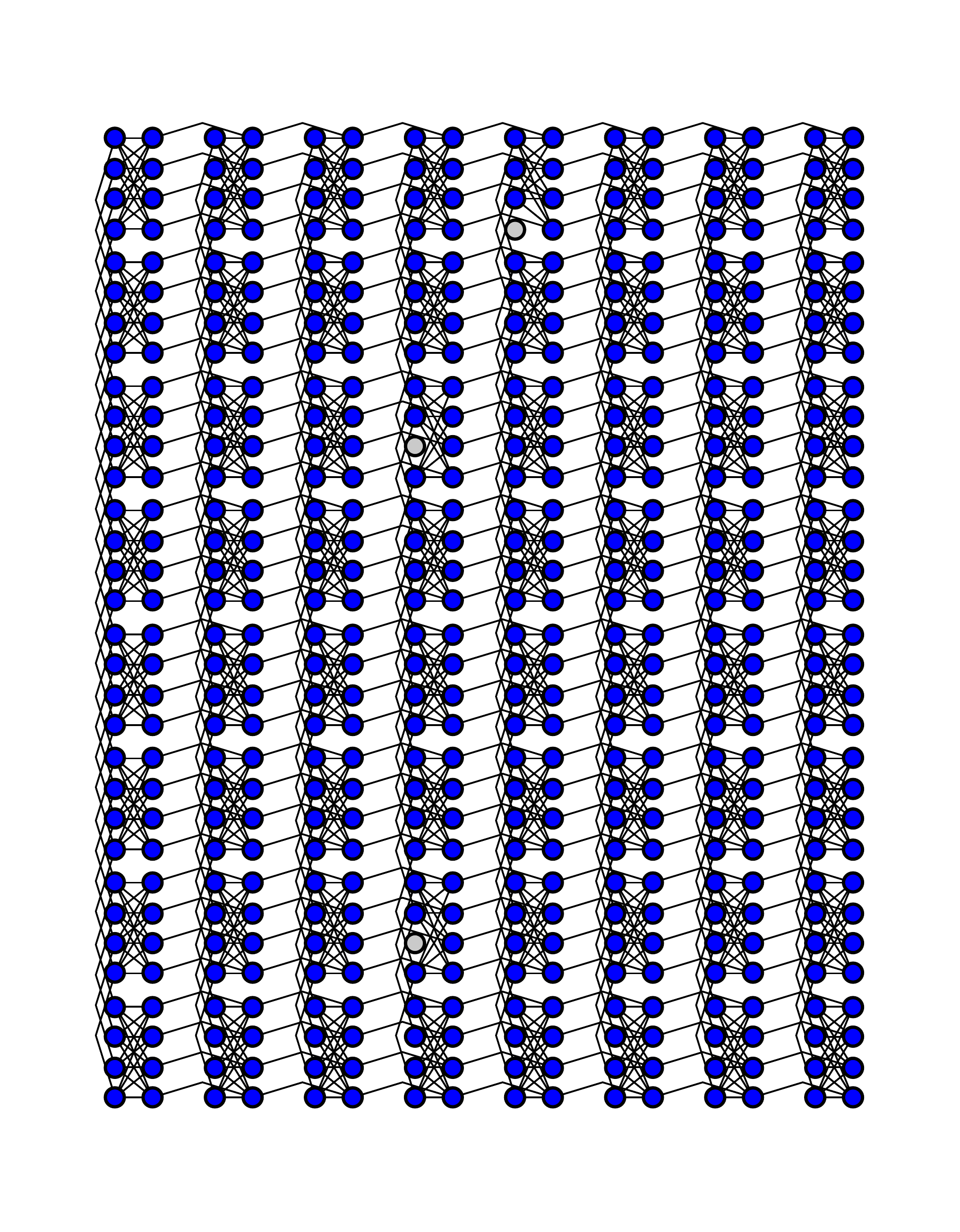}
\caption{\scriptsize{\textbf{Device architecture and qubit connectivity:}. The array of superconducting quantum bits is arranged in $8\times 8$ unit cells that consist of 8 quantum bits each. Within a unit cell, each of the 4 qubits in the left-hand partition (LHP) connects to all 4 qubits in the right-hand partition (RHP), and vice versa. A qubit in the LHP (RHP) also connects to the corresponding qubit in the LHP (RHP) of the units cells above and below (to the left and right of) it. Edges between qubits represent couplers with programmable coupling strengths. Blue qubits indicate the 509 usable qubits, while grey qubits indicate the three unavailable ones out of the 512 qubit array.}}
\label{fig:chimera}
\end{figure}

The quantum hardware employed consists of 64 units of a recently characterized eight-qubit unit cell~\cite{harris2010,johnson_quantum_2011}. Post-fabrication characterization determined that only 509 qubits out of the 512 qubit array can be reliably used for computation (see Fig.~\ref{fig:chimera}). The array of coupled superconducting flux qubits is, effectively, an artificial Ising spin system with programmable spin-spin couplings and transverse magnetic fields. It is designed to solve instances of the following (NP-hard~\cite{Barahona1982}) classical optimization problem: Given a set of local longitudinal fields $\{h_i\}$ and an interaction matrix $\{J_{ij}\}$, find the assignment $\mathbf{s^*} = s^*_1 s^*_2 \cdots s^*_N$, that minimizes the objective function $E(\mathbf{s})$, where,
\begin{equation}\label{eq:QUBO}
E(\mathbf{s})  = \sum_{1 \le i \le N} h_{i} s_i  + \sum_{1 \le i<j\le N} J_{ij} s_{i} s_{j},
\end{equation}
$\abs{h_i} \le 2$, $\abs{J_{ij}} \le 1$, and $s_i \in \{+1,-1\}$.

Finding the optimal $\mathbf{s^*}$ is equivalent to finding the ground state of the corresponding Ising classical Hamiltonian,
\begin{equation}\label{h-ising}
H_{p}  =  \sum^N_{1 \le i \le N} h_{i}\sigma_{i}^{z}  + \sum^N_{1 \le i<j\le N} J_{ij}\sigma_{i}^{z} \sigma_{j}^{z}
\end{equation}
where $\sigma_{i}^{z}$ are Pauli matrices acting on the $i$th spin.

Experimentally, the time-dependent quantum Hamiltonian implemented in the superconducting-qubit array is given by,
\begin{equation}\label{h-AQO}
H(\tau)  = A(\tau) H_b + B(\tau) H_p, \quad \quad \tau= t/t_{a},
\end{equation}
with $H_b  = - \sum_i \sigma^{x}_{i}$ responsible for quantum tunneling among the localized classical states, which correspond to the eigenstates of $H_p$ (the computational basis). The time-dependent functions $A(\tau)$ and $B(\tau)$ are such that $A(0) \gg B(0)$ and $A(1) \ll B(1)$; in Fig.~\ref{fig:cb21f6}, we plot these functions as implemented in the experiment. $t_{a}$ denotes the time elapsed between the preparation of the initial state and the measurement, referred to hereafter as the \textit{annealing time}.

QA exploits the adiabatic theorem of quantum mechanics, which states that a quantum system initialized in the ground state of a time-dependent Hamiltonian remains in the instantaneous ground state, as long as it is driven sufficiently slowly. Since the ground state of $H_p$ encodes the solution to the optimization problem, the idea behind QA is to adiabatically prepare this ground state by initializing the quantum system in the easy-to-prepare ground state of $H_b$, which corresponds to a superposition of all $2^N$ states of the computational basis. The system is driven slowly to the problem Hamiltonian, $H(\tau=1) \approx H_p$.  

Determining the optimum value of $t_a$ is an important and non-trivial problem in itself. In
principle, the adiabatic theorem states that over sufficient adiabatic time
$t_a$, the state $\ket{\psi(t_a)}$ will converge to the solution of the problem
$\ket{\psi_{g} (t_a)}$. Notice that the parameter $t_a$ determines the rate at which $H(t)$ varies.
Following the notation from Farhi {\it et al}~\cite{farhi2000}, consider $H(t)=
\tilde{H} (t/ t_a) = \tilde{H}(\tau)$, with instantaneous values of
$\tilde{H}(\tau)$ defined by
\begin{equation}
 \tilde{H}(\tau)\ket{l;\tau}=E_{l}(\tau)\ket{l;\tau}
\end{equation}
with
\begin{equation}
 E_{0}(\tau) \le E_{1}(\tau) \le \cdots \le E_{M-1}(\tau)
\end{equation}
where $M$ is the dimension of the Hilbert space, e.g., for the case of $N$ qubits $M=2^N$. According to the adiabatic theorem, if the gap between the two lowest levels, $E_{1}(\tau) - E_{0}(\tau)$, is greater than zero for all $0 \le \tau \le 1$, and taking
\begin{equation}
 t_a \gg \frac{\varepsilon}{g_{min}^{2}}
\end{equation}
with the minimum gap, $g_{min}^{2}$, defined by
\begin{equation}
 g_{min}= \min_{0 \le \tau \le 1} (E_{1}(\tau) - E_{0}(\tau)),
\end{equation}
and $\varepsilon$ given by
\begin{equation}
 \varepsilon = \max_{0 \le \tau \le 1} \vert \braket{l=1;\tau}{\frac{d\tilde{H}}{d\tau} \vert l=0;\tau} \mid,
\end{equation}
then we can make
\begin{equation}
 \vert \braket{l=0;\tau=1}{\psi(t_a)} \vert
\end{equation}
arbitrarily close to 1. In other words, the existence of a nonzero gap guarantees that $\ket{\psi (t)}$ remains very close to the ground state of $H(t)$ for all $0 \le t \le t_a$, if $t_a$ is sufficiently large.

The adiabatic condition presented above applies to zero-temperature noise-free (i.e., 100\% quantum coherent) conditions, where the dynamics are well-described by $H(t)$ in Eq.~\ref{h-AQO} and the scaling of $t_a$ as a function of the problem size determines the complexity (and usefulness) of the quantum algorithm. In a realistic experimental implementation, the quantum processor will operate at a finite temperature, and in addition to thermal fluctuations, other types of noise are unavoidable, leading to dissipation processes not captured in $H(t)$; deviations from adiabaticity affecting the performance of the quantum algorithm seem to be a delicate balance between the quantum coherence effects and the interaction with the environment, responsible for issues like thermal excitation (relaxation) processes out of (into) the ground state~\cite{AlbashNJP2012,PerdomoOrtiz2012_LPF}. For example, as shown in Fig.~\ref{fig:cb21f6}b, contrary to what is expected from the adiabatic condition, longer annealing times $t_a$ do not necessarily imply a monotonic enhancement in the success probability. To the best of our knowledge this question related to the scaling of $t_a$ in an noisy environment is still largely unexplored. From an experimental standpoint, the main limitation is the limited size of the available quantum devices.

The first challenge of the experimental implementation is to map the computational problem of interest into the binary quadratic expression (Eq.~\ref{h-ising}), which we outline next. In the next subsection we describe the computational problem and in the following we present its mapping to QUBO.

\subsection{Diagnosing multiple faults in electrical power networks}\label{subsec:DMF}

\begin{figure*}
\centering
\includegraphics[width=0.95\textwidth]{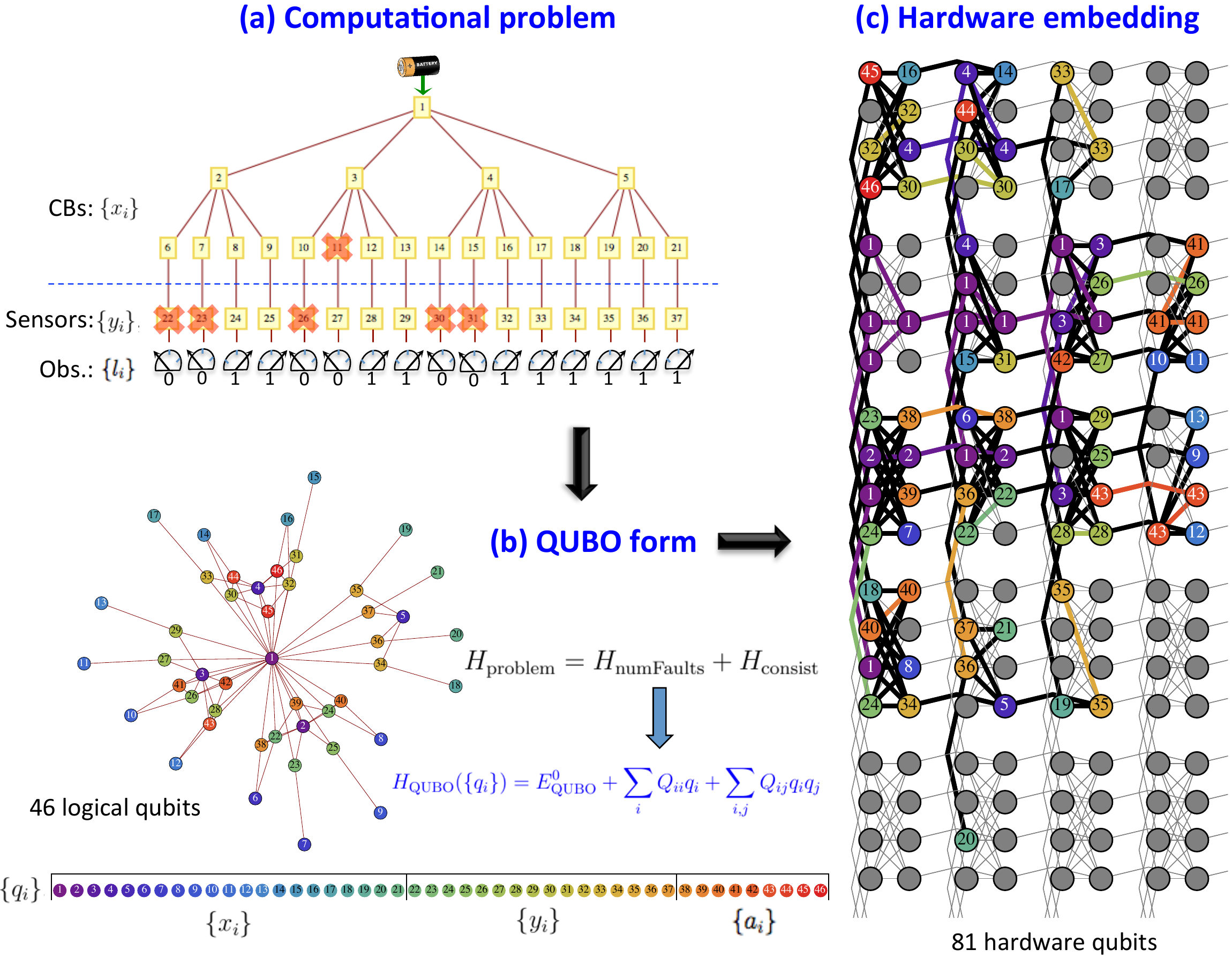}
\caption{\scriptsize{General scheme of an experimental realization for the diagnosis of multiple faults with a quantum annealing device. (a) A possible realization of the diagnosis of multiple faults problem mentioned in Sec.~\ref{subsec:DMF}. This particular realization consists of an EPS network with one power source, 21 circuit breakers and 16 ammeters or sensors. To capture the uncertainty in the observations (ammeter readouts), we introduce a binary variable per ammeter (leaves of the tree). Like the case of the binary variable for the CBs, these variables will be diagnosed with a zero (one) when the sensor is faulty (healthy). The orange crosses indicate faulty electrical components ($x_i=0)$. In this particular realization of six faults, this plausible explanation places one of the faults on the CBs and the remaining five on the ammeters. This is one of the $2^6$ six-fault explanations that are equally likely in this problem instance. (b) The problem needs to be mapped to a quadratic unconstrained binary optimization (QUBO) expression as described in Sec.~\ref{subsec:mapping2qubo}. The \textit{QUBO graph} is a representation of its QUBO energy expression, $H_{\text{QUBO}}$, where each $Q_{ij}$ coupling among two logical qubits in Eq.~\ref{eq:Hqubo} is represented as edges in the graph. (c) The subsequent embedding into the hardware architecture usually requires more variables, since some logical qubits are represented by several physical qubits (represented here as nodes in the graph) due to the sparse connectivity of the hardware graph (notice that each node in the hardware architecture is connected to at most six neighbors). In this 21 CB problem, 81 physical qubits are needed to implement the QUBO graph with 46 logical variables. More details on the minor embedding problem for the D-wave device can be found in Ref.~\onlinecite{Cai-14}}}
\label{fig:diagnosis}
\end{figure*}

One of the most general and ubiquitous systems is an electrical power-distribution structure made of power sources, a distribution network, and power sinks (loads). The distribution network structure is typically made up of circuit breaker (CB) components, represented as nodes on a graph, and wires between them, represented as lines in a graph, as illustrated in Fig.~\ref{fig:diagnosis}a. Typically sensors are placed at different locations that measure certain properties of the system at that location. Some properties that can be sensed include current, voltage, temperature etc. Because sensors are also physical devices and because of inherent uncertainty in data acquisition, the readouts from sensors can be noisy and may be incorrect if the sensors themselves have failed in some way.

Consider for example the network structure shown in Fig.~\ref{fig:diagnosis}a, consisting of a quaternary tree powered by a single battery source and with electrical current flowing down through intermediate CB nodes to the leaf sensor nodes. Suppose that the CBs and the ammeters might be faulty - affecting the propagation of current through the network and introducing uncertainty in the observation readouts, respectively. To simply the discussion, we assume in this report that the failure probability of the CBs is the same as the one for the ammeters. Under this assumption, the goal of the multiple-fault-diagnosis problem is to find the minimal number of faults, or broken electrical components (CBs and/or ammeters), that are consistent with the readout from the circuit. It is assumed here that power is generated only from this single node, and therefore will cause the ammeters on the other side to signal the flow of current, barring any faults in the connection point from starting point to these sensors, unless the most probable explanation (minimum number of faults) include the failure of the ammeters themselves. Note that our approach explained below can be generalized to more complex problems like multiple power sources, arbitrary tree/graph structures, sensors at intermediate locations, etc.

\subsection{Mapping to QUBO}~\label{subsec:mapping2qubo}
As explained in the introduction to the section, we need to convert the circuit layout for the electrical network into a mathematical form that is amenable to the quantum computer's problem-solving capabilities. As mentioned before, this mathematical form is a QUBO. Our problem is naturally expressed in terms of 0's representing faulty components and 1's representing healthy components. Below, we demonstrate the procedure of taking this problem and converting it into QUBO form so that it can be solved by the D-Wave computer.

As shown in Fig.~\ref{fig:diagnosis}a, there are two types of components: i) CBs which in their healthy mode allow the flow of current, and are illustrated as the nodes of the quaternary tree. We will denote them by the set of binary variables $\{x_i\}$, with $x_i=1$ ($x_i=0$) corresponding to CB $i$ in a healthy (faulty) state. ii) The other type of component is the sensor or ammeter, which is not only another type of electrical component that could in principle malfunction, but also forms part of the measurements or observations from which one is asked to perform the diagnosis of the electrical network. Therefore, for each ammeter we will have an observation parameter and a sensor-status variable indicating its healthy or faulty status. The observations or readouts are given as part of the definition of the problem and given as input parameters. We will denote this set of binary parameters $\{l_i\}$, with $l_i=1$ ($l_i=0$) if the $i$-th ammeter is showing a HIGH (LOW) readout. Similar to the $\{x_i\}$ variables for the CBs, the uncertainty in the ammeter readouts is introduced by assigning to them a set of binary variables, $\{y_i\}$, with $y_i=1$ ($y_i=0$) corresponding to ammeters $i$ in a healthy (faulty) state.

The goal is to find the minimum number of faults in the electrical components, either on CBs and/or ammeters, consistent with the circuit layout and the readouts. We solve this as a minimization problem over the pseudo-boolean function $H_{\text{problem}}(\{x_i\},\{y_i\};\{l_i\})$. After $H_{\text{problem}}$ is transformed into its QUBO form, we can subsequently use the quantum computer to find the assignment for each of the $\{x_i\}$ and $\{y_i\}$. 

The construction of the pseudo-boolean function contains two contributions:
\begin{equation}
\label{eq:Hproblem}
H_{\text{problem}} = H_{\text{numFaults}} + H_{\text{consist}}.
\end{equation}

$H_{\text{consist}}$ is constructed such that it is zero whenever the prediction from the assignment of all the $\{x_i\}$ and $\{y_i\}$ is consistent with the readouts $\{l_i\}$ from the ammeters, and greater than zero when the readouts and the prediction, given the assignments of the $\{x_i\}$ and $\{y_i\}$, do not match. Consider the set $P_i$ as the set of CB indices in the path from the root node (CB 1) where power is inputted, all the way to the CB connected to $i$-th ammeter. For example, for the network in Fig.~\ref{fig:diagnosis}a,  $P_1 = \{1,2,6\}$, $P_2 = \{1,2,7\}$, $P_3 = \{1,2,8\}$, $\cdots$, and $P_{16} = \{1,5,21\}$. If we denote the number of paths as $n_{\text{paths}}$ (in this network it equals the number of ammeters), one can construct $H_{\text{consist}}$ as, 
\begin{equation}\label{eq:Hconsist}
H_{\text{consist}} = \lambda_{\text{path}} \sum_{i=1}^{n_{\text{paths}}}  y_i g_i, \quad   f_{i}(\{x_j\}_{j \in P_i}) = \prod_{j\in P_i} x_j,
\end{equation}
with $g_i = l_i + f_{i}-2f_i l_i$, a binary function with $g_i=0$ when the prediction $f_i$, based only on the CB statuses in the path $P_i$, is consistent with the readouts $l_i$, and $g_i=1$ when the prediction and the readout are in disagreement. In other words, $g_i = \textsc{xor}(f_i,l_i)$

$H_{\text{numFaults}}$ is proportional to the number of faults (whenever $x_i = 0$ or $y_i = 0$) in the electrical network,
\begin{equation}
\label{eq:Hfaults}
H_{\text{numFaults}} = \lambda^{\text{CB}}_{\text{faults}} \sum_{i=1}^{n_{\text{CB}}} (1-x_i)+ \lambda^{\text{sensor}}_{\text{faults}} \sum_{i=1}^{n_{\text{sensor}}} (1-y_i),
\end{equation}
and when combined with $H_{\text{consist}}$, as written in Eq.~\ref{eq:Hproblem}, defines the problem energy function to be minimized by favoring the minimal set of faulty components that are simultaneously consistent with the observations measured in the outermost sensors.

There is an important difference between the implementation of the search with a quantum algorithm when compared to an implementation with a conventional classical computer. Doing \textsc{if} loops and checking whether or not a certain solution is consistent with the observations is an easy task on a conventional computer, since one can interrupt the process at any time to accept or to reject a solution candidate to the problem. In a quantum device, this checking procedure of measuring whether the state is in accordance with the observations, can only be done once, which happens  at the end of the quantum annealing schedule, since according to quantum mechanics any measurement of the state collapses the state to a classical outcome (one of the solutions); therefore, if the result is not the desired solution, then one needs to restart another optimization cycle from scratch. This is the main reason why both terms, $H_{\text{consist}}$ and $H_{\text{faults}}$, need to be part of the same cost energy function to be optimized, as expressed in Eq.~\ref{eq:Hproblem}. Since these quantum algorithms attempt to find the lowest energy solutions, the method we use to cope with this ``no-restart" issue is to set the penalty $\lambda_{\text{path}}$ high enough to avoid considering solutions that are inconsistent with the observations. Ideally setting $\lambda_{\text{path}} \rightarrow \infty$ would solve this issue, but due to the limited range of values that can be programmed in the device, this is not possible. At least one should find a lower bound for this penalty such that it guarantees that the desired optimal solution (minimum number of faults satisfying the observation constraints) always has a lower value of $E_{\text{problem}}$ than any assignment violating the observation constraints ($H_{\text{consist}}>0$). For example, for the case of the mapping with uncertainty in the sensors described above in Eq.~\ref{eq:Hconsist} and Eq.~\ref{eq:Hfaults}, one can prove that as long as $\lambda_{\text{path}} >\lambda_{\text{faults}}$, the optimal solution corresponding to the minimal number of faults will always have an $E_{\text{problem}}$ lower than any assignment of $\{x_i,y_i\}$ that is not in agreement with the observation. This is a consequence of the tree structure of the problem and the fact that there can only be penalties due to inconsistencies in the observations if the $i$-th ammeter is healthy ($y_i =1$), since whenever $y_i =0$, the contribution to $H_{\text{consist}}$ would be zero.  It can be proved by considering that the energy of any inconsistent configuration incurring a penalty via $H_{\text{consist}}$ can always be lowered by flipping the sensor in its path; this will make the assignment a consistent one. Flipping this sensor to a faulty state will raise the energy by $\lambda_{\text{fault}}$, but at the same time the energy is lowered by a greater amount $\lambda_{\text{path}}$, as long as $\lambda_{\text{path}} >\lambda_{\text{faults}}$. In the example presented here, we set $\lambda_{\text{path}} = 3$ and $\lambda_{\text{faults}} =1$.

For the 21 CB, 6 fault problem, shown in Fig.~\ref{fig:diagnosis}a where $3$ of the $4$ outermost sets of $4$ sensors contain two $l_i=0$, the solution must contain a composite of $6$ components set to $0$. These $6$ total failures can be any combination of sensors equal to $0$ or outermost CB’s equal to $0$, as long as the total number summed amounts to $6$ and each failure corresponds to exactly one of the branches with $l_i=0$. 

Notice the pseudo-boolean $H_{\text{consist}}$ is a high-degree polynomial, and for this particular network, the order of the polynomial is related to the depth of the tree. We can reduce the degree of the polynomial to a quadratic expression, $H_{\text{QUBO}}$, with the overhead of adding more binary variables, while conserving the global minimum of the original pseudo-boolean function, $H(\{x_i\},\{y_i\};\{l_i\})$. Further details on the new techniques used for this reduction are provided in the appendix.

Assuming it requires $n_A$ ancilla variables $\{a_i\}$ to reduce the high-degree polynomial to the quadratic expression, we can relabel the CB, sensor, and ancilla variables, $\{x_i\}$, $\{y_i\}$, and $\{a_i\}$, respectively, into a new set of binary variables $\{q_i\}$ for $i = 1,2,3, \cdots, n_{l}$, with $n_{l} = n_{\text{CB}}+ n_{\text{sensor}}+n_{A}$ as the total number of logical qubits. The final quadratic cost function to be minimized can be written as
\begin{equation}\label{eq:Hqubo}
\begin{split}
H_{\text{QUBO}}(\{q_i\}) &= E^0_{\text{QUBO}} + \sum_{i,j}  Q_{ij} q_i q_j \\&= E^0_{\text{QUBO}} + \mathbf{q}^T \cdot \mathbf{Q} \cdot \mathbf{q}.
\end{split}
\end{equation}
As shown in Fig.~\ref{fig:diagnosis}, this final expression can be represented as a graph with the number of vertices equal to the number of logical qubits $n_{l}$ corresponding to the set of variables $\{q_i\}$. In this representation, $Q_{ii}$ can be seen as the weights on the vertices, while $Q_{ij}$ are the weights for the edges representing the couplings between variables $i$ and $j$ (see Fig.~\ref{fig:diagnosis}). Notice that since $q^2_i = q_i$, the expression $\mathbf{q}^T \cdot \mathbf{Q} \cdot \mathbf{q}$ contains both linear terms $Q_{ii}$, and quadratic term, $Q_{ij}$, when $i \neq j$.   $E^0_{\text{QUBO}}$ corresponds to the constant independent term.

A simple mathematical transformation of the form $q_i = (s_i+1)/2$ allows to rewrite $H_{\text{QUBO}}$ in the equivalent problem:
\begin{equation}\label{eq:Hising}
\begin{split}
H_{\text{Ising}}(\{s_i\}) &= E^0_{\text{Ising}} + \sum_{i}  h_{i} s_i + \sum_{i>j}  J_{ij} s_i s_j \\&= E^0_{\text{Ising}} + \mathbf{h} \cdot  \mathbf{s} + \mathbf{s}^T \cdot \mathbf{J} \cdot \mathbf{s}.
\end{split}
\end{equation}
where one is looking for a minimization over the new set of of so called \textit{spin} variables $\mathbf{s} = \{s_1, s_2, \cdots, s_{n_l}\}$, where now each readout of $s_i = -1 (+1)$ correspond to the assignments $q_i  = 0 (1).$

\section{Results and Discussion}\label{sec:ResultsDiscussion}

As shown in Fig.~\ref{fig:diagnosis}c, the quantum device has a well-defined architecture, which in turn determines the possible connectivity among the qubits. Due to the sparse connectivity of the hardware graph, it is clear that variables in dense or fully connected graphs cannot be mapped one-to-one to qubits in the device, resulting in an overhead of the number of qubits needed for the experimental realization. This overhead is necessary because some logical qubits in $\{q_i\}$ need to be replicated among several physical (hardware) qubits in order to fulfill the connectivity requirements in the original denser graph. The problem of finding this correspondence between logical and physical qubits is what we call \textit{the embedding problem}. A more detailed description of this problem can be found elsewhere~\cite{Choi2008,Choi2011,Klymko2013,Cai-14}. For the purpose of our discussion, we use the embedding algorithm~\cite{Cai-14} as implemented in D-wave's Application Programming Interface, which allows us to obtain an assignment of the $n_{l}$ logical qubits to the $n_p$ physical (hardware) qubits as the final step before the processor can be programmed to solve the optimization problem of interest. Further details on how to set the parameters of the embedded Hamiltonian can be found in Ref.~\onlinecite{APO2014_DWtuning}.
To study the number of qubits required for the experimental implementation of the diagnosis of multiple faults problem described in Sec.~\ref{subsec:DMF}, we embedded EPS networks similar to those described in Fig.~\ref{fig:diagnosis}. In Table~\ref{table:embedding_HyDE}, we present the scaling of the number of resources, $n_p$ (column 4), needed in the hardware for a few select problem instances of different tree size and various numbers of faults.

\begin{table*}[!t]
\renewcommand{\arraystretch}{1.3}
\caption{Embedding for quaternary tree networks in the case of uncertainty in the sensor readouts and running time to solution with the classical solver HyDE, when available. As described in the text, $n_{l}$ and $n_p$ correspond to the number of logical and physical qubits, respectively.}
\label{table:embedding_HyDE}
\centering
\begin{tabular}{c|c|c|c|c|c|c}
\hline
$n_{\text{CB}}$ & $n_{\text{sensors}}$ & $n_{l}$ & $n_p$ & $n_{\text{faults}}$ & Problem instance, $\{l_i\}$ & HyDE time (s)\\
\hline
5 & 4 & 12 & 17 & 2 & $\{0001\}$ \\ 
21 & 16 & 42 & 68 & 3 & $\{0101,0111,1111,1111\}$ & 0.164 \\
21 & 16 & 45 & 73 & 4 & $\{0101,0101,1111,1111\}$ & 3.129\\
21 & 16 & 46 & 78 & 5 & $\{0101,0101,0111,1111\}$ & 516.0\\
21 & 16 & 46 & 81 & 6 & $\{0101,0101,0101,1111\}$ & out of memory\\ 
21 & 16 & 48 & 90 & 7 & $\{0101,0101,0101,0111\}$ & out of memory\\
21 & 16 & 49 & 96 & 8 & $\{0101,0101,0101,0101\}$ & out of memory\\
85 & 64 & 165 & 340 & 6 & $\{01 1 1, 1 1 1 1, 0 1 1 1, 1 1 1 1, 0 1 1 1, 1 1 1 1, 1 0 1 1, 1 1 1 1,$ \\
 &  &  &  & & $0 1 1 1, 1 1 1 1, 1 1 1 1, 1 1 1 1, 0 1 1 1, 1 1 1 1, 1 1 1 1, 1 1 1 1\}$ \\
\hline\hline
\end{tabular}
\end{table*}

In the last column of Table~\ref{table:embedding_HyDE}, we report the running time it took for HyDE to solve the same problem instances. HyDE is a general purpose model-based diagnosis engine. HyDE uses models of nominal and faulty behavior of the system being diagnosed to make predictions about the expected trajectory of the system. This can be compared against the actual system trajectory as observed through sensors placed in the system to detect any discrepancies. HyDE then uses the detected discrepancies to perform a heuristic search over the fault space to select most likely candidates. One or more candidates consistent with the sensor observations are reported as the likely faults.

HyDE supports a wide variety of models and sensor types. The models can at different levels of abstractions, qualitative or quantitative in nature, be static or dynamic (differential equations), and include uncertainties. Sensors can be boolean, enumerations, real-valued or interval-valued. While this allows HyDE to be used in a wide variety of domains and problems, HyDE cannot take advantage of special characteristics of systems and domains unless explicitly modeled.

Figure~\ref{fig:diagnosis}a shows the HyDE model of electrical network described in the previous section. HyDE uses the model to predict the expected readings of the ammeters. This is compared against the actual ammeter readings to determine discrepancies. Since the ammeters themselves can be faulty, for each discrepant ammeter the ammeter itself is faulty or there is at least one CB upstream that is faulty. However HyDE cannot necessarily eliminate all CBs connected to non-discrepant ammeters as candidates. For example it is possible that CB and a downstream ammeter might be both faulty resulting in the ammeter not being characterized as discrepant. HyDE uses the above heuristics to reduce the search space but in the worst case it still turns out to be exponential search. For example when looking for a 5 fault candidate, HyDE still has to eliminate all 4 fault candidates (looking for the smallest diagnosis). Some of these can be eliminated by the heuristics but the rest have to still be tested by HyDE.

The last column of Table~\ref{table:embedding_HyDE} shows that problems can become intractable very quickly as one increases the number of faults and illustrates this exponential growth in the search space as a function of the number of possible faults, for instances with a fixed number of electrical components. For the problems studied, which were run on an Intel(R) Code(TM) i7-2720QM CPU at 2.20 GHZ and 8GB of RAM workstation, the computer exited without providing an answer due to memory insufficiency when there were 6 or more faults. Even though we studied all the cases reported in the table, and many other benchmark problems with networks of binary and quaternary trees of different sizes, we focused on the instance of 21 CBs, 16 sensors, and 6 faults since this was the first small scenario where HyDE, as an exact solver, failed to provide a diagnosis when run on a conventional workstation. For this problem, the number of logical binary variables, $n_l$, for each problem instance can be split among the 37 binary variables providing the relevant information about the 21 CBs and 16 sensors, and the remaining ancilla variables needed for the final construction of $H_{\text{QUBO}}$, as described at the end of Sec.~\ref{subsec:mapping2qubo}.

\begin{figure*}
\centering
\includegraphics[width=0.95\textwidth]{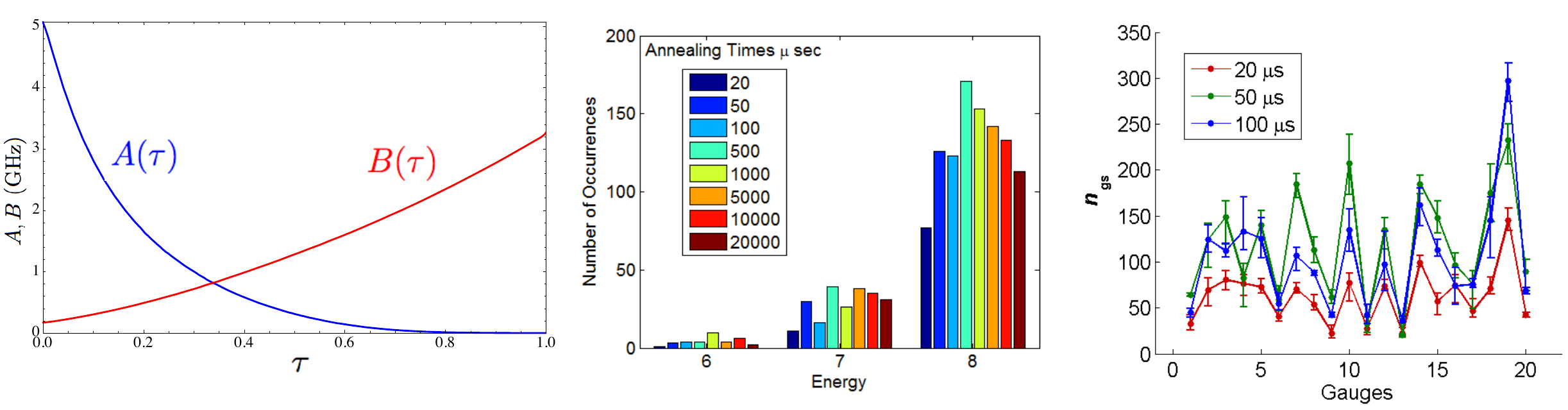}
\caption{\scriptsize{(Left) Time-dependence of the $A(\tau)$ and $B(\tau)$ functions, where $\tau=t/t_{a}$. (Center) Experimental results from running the quantum annealing algorithm at different annealing times. For each annealing time, the number of occurrences are considered over 50,000 cycles of the quantum annealing schedule. We report the three lowest energy configurations, with $E=6$ corresponding to the desired 6-fault solutions. Note that as a bonus, when compared to classical diagnostics tools such as HyDE that, by default, only report the optimal solution, the quantum algorithm also reports solutions with $E=7$ and $E=8$ (seven faults and eight faults, respectively), corresponding to the next best set of candidates that also explain the observations. This might be useful in cases where the most probable explanation does not correspond to the actual state of the system. (Right) Experimental results at three different annealing times for 20 different random gauges, where gauge \#1 represents of case of no gauge transformation being applied. The number of occurrences in the ground state with $E=6$, $n_{gs}$, are based on the average of three repetitions with 100,000 readouts each. The data in the plot shows this mean value, while the error bars correspond to the lowest and highest of these three outcomes for the purpose of showing the fluctuations from the quantum device over these 100,000 sets.}}
\label{fig:cb21f6}
\end{figure*}

Quantum annealing is designed to sample the low energy (low cost) solutions of the objective cost function $H_{\text{QUBO}}$, but the algorithm does it in a probabilistic manner. Therefore, the strategy for finding the solution is to run the algorithm at least a certain number of repetitions, $R_P$, large enough to find the desired optimal solution with a probability $P$ close to 1.0. Let's define by $R_{.99}$ the number of repetitions needed to find the optimal solution with a certainty of 99\%. $R_{.99}$ clearly depends on the probability $p_s$ of measuring this desired optimal solution after a single repetition, which in turn depends on the the annealing time, $t_a$, specified as input for each repetition or cycle, i.e., $p_s = p_s(t_a)$. Suppose you request a number of readouts $N_r$ from the quantum processor. Then $p_s(t_a)$ can be estimated as the ratio $n_{gs}/N_r$, with $n_{gs}$ denoting the number of occurrences of the ground states (corresponding to the desired solutions), at the specified $t_a$. For example, the success probability $p_s(t_a)$ can be determined from the histograms in Fig.~\ref{fig:cb21f6}a by dividing the number of occurrences of the optimal solutions with $E=6$ by the total number of runs $N_r$, set to 50,000 in this particular instance in order to obtain at least some solutions.  

The number of repetitions needed to reach $R_P$ can be estimated as follows. The probability of not obtaining the lowest configuration after $R_P$ runs is $(1-p_s)^{R_P}$. Therefore, the probability of measuring the ground state at least once in these $R_P$ experimental runs is given by $P = 1-(1-p_s)^{R_P}$. Thus, the number of repetitions $R$ one needs to run the quantum annealing algorithm to find the lowest energy solution with at least a probability $P$ is determined by

\begin{equation}\label{eq:repetitions}
R_P(t_a) = \ceil*[\big]{\frac{log[1-P]}{log[1-p_s(t_a)]}}
\end{equation} 

Let's define the time-to-solution $t_{QA}$ as the time required in the quantum processor to find the solution to the problem with a certainty of 99\%, each cycle requiring an annealing time $t_a$. Then, by estimating $p_s(t_a)$, $t_{QA}(t_a)$ can be calculated as
\begin{equation}\label{eq:totaltime}
t_{QA}(t_a) = R_{.99}(t_a) t_a.
\end{equation} 

Estimation of the time-to-solution, $t_{QA}$, based on $R_{.99}$ has proven to be useful in scaling studies comparing the D-wave processor with state-of-the-start processors~\cite{boixo_evidence_2014,Ronnow2014_quantumspeedup}. Since estimating this time requires finding the solution to the real-world problem a sizable number of times $n_{gs}$ in calculating $p_s$, one might ask the question: Is there any purpose to estimating these values given that, from an application/real-world application perspective, finding the solution just once is all that is desired? We think there are several good reasons to form these experimental estimates:
\begin{enumerate}
\item To compare the difficulty among problems in different application domains.
\item To compare between different realizations of the same problem, for example, under different gauge realizations as shown in Fig.~\ref{fig:cb21f6}b and discussed below.
\item To obtain the scaling of the algorithm runtime as a funtion of the problem size for families of problems within a certain application
\end{enumerate}

Since the work presented here focuses on only one application, we do not go into depth on argument \#1, but it might be good to point out that as more applications are implemented in quantum annealers (e.g. lattice protein folding~\cite{PerdomoOrtiz2012_LPF}, operational planning and scheduling~\cite{planningquantum}, solar-flares~\cite{ogorman2014_bayesnet}), it would be interesting to compare the performance of the machine and the hardness of these applications as a function of problem size. For example, it is interesting to note that the relatively small problem instances studied here, with only 81 physical qubits, seem to be at least as hard as the average problem with 503 qubits for random instances presented in Ref.~\onlinecite{Ronnow2014_quantumspeedup}. In that case, most of the instances were solved at least once in the 1000 runs performed ($p_s > 1/1000$) at $20 \mu$s, while the six-fault scenario presented in Fig.~\ref{fig:diagnosis} was solved only 33 times in 100,000 runs for the case of no gauge ($p_s < 1/1000$), and even at the best gauge the number of occurrence was still a modest 145 in 100,000 runs on average ($p_s = 1.45/1000$). The hardness of this small problem with 81 qubits might not be necessarily related to its intrinsic hardness compared to the 503 qubit instances in Ref.~\onlinecite{Ronnow2014_quantumspeedup}; we believe it is related to calibration errors in the D-wave machine~\cite{harrisPRB2010} that can go up to 5\% in the specification of the programmable parameters, $h_i$'s and $J_{ij}$'s. This will significantly affect real-world motivated applications since they contain problem-defined real constraints, i.e., the $h_i$'s and $J_{ij}$'s to be programmed assume real values instead of the random but well-defined integer values used in other benchmark studies, e.g. only $J_{ij} =\pm 1$. These calibration errors will lead to the specification of a slightly different problem that may return a solution different from the desired one. Improving the precision of the D-wave device should be one of the top priorities in enabling this state-of-the-art device to solve real-world problems.

It is known that the effect of calibration errors on the D-wave machine can be significantly reduced by repeating the annealing runs for several different realizations of the same problem under different \textit{gauge} transformations.~\cite{boixo_evidence_2014,Ronnow2014_quantumspeedup}. After the embedding procedure, the experimentally-realized Hamiltonian resembles that of $H_{\text{Ising}}$ in Eq.~\ref{eq:Hising} (over a set of $n_p$ hardware qubits rather than the original $n_l$ qubits from the unembedded graph). A gauge corresponds to a mathematical transformation specified by a vector $\mathbf{a} = \{a_1, a_2, \cdots, a_{n_p}\}$, with $a_i = \pm 1$, transforming each $s_i \rightarrow a_i s_i$ in Eq.~\ref{eq:Hising}. Note that this transformation does not change the energy landscape where the optimization is performed, since the original assignment can be trivially obtained by reversing the sign of the value measured for $s_i$. According to Eq.~\ref{eq:Hising}, the net effect of the gauge is to transform $h \rightarrow a_i h_i$ and $J_{ij} \rightarrow a_i a_j J_{ij}$. Although this transformation should not affect the performance of algorithms implemented in classical computers, the experimental implementation in the D-wave quantum device breaks this gauge symmetry due to calibration and precision limitations in the device~\cite{harrisPRB2010}, i.e., for example the precision of setting positive values of $h_i$ or $J_{ij}$ could be different than the precision for setting negative values. 

Fig.~\ref{fig:cb21f6}b illustrates this point. We randomly generated 20 gauges and performed the runs using these gauges. The runs with the different gauges resulted in varying occurrences in the ground states, $n_{gs}$. Since the number of ground state occurrences is proportional to the success probability of finding the solution, $p_s(t_a)$, the results indicate that the gauges have a significant impact on the performance of the machine. Out of this 20 gauge set, gauge \#19 gave the largest enhancement in the number of ground states over the ungauged configuration (corresponding to gauge \#1). These enhancements, represented by $n_{gs, g19}/n_{gs,g1}$, correspond to 4.4, 3.6, and 6.6, as a function of the different annealing times 20 $\mu$s, 50 $\mu$s, and 100 $\mu$s, respectively.

The results in Table~\ref{table:results} show the values of $R_{.99}$ and the corresponding time, $t_{QA}$, in the quantum processor for different annealing times. The calculations of the average values reported are simply the average over the 20 predicted values of $R_{.99}$ and of $t_{QA}$ for each one of the gauges, which is a simpler procedure than the one proposed in Ref.~\onlinecite{boixo_evidence_2014}. The mean is an assessment of the expected number of repetitions (or the expected processor time) if one were to select a random gauge. The values obtained using the best gauge are also reported. There is no way \textit{a priori} to select what the top gauge would be, and it is an impractical task to do an exhaustive search over the $2^{81}$ possible gauge realizations in this 81 qubit experiment. Given the significant enhancement in perfomance, we address this issue in a follow-up publication, where we propose a method to select the best gauges out of a pool of randomly generated gauges, similar to the ones generated here.~\cite{APO2014_DWtuning}

\begin{table}[!t]
\renewcommand{\arraystretch}{1.3}
\caption{Repetitions $R_{.99}$ needed to find the solution to the computational problem with a certainty of 99\% and effective runtime $t_{QA}$ for the 21 CB, 6 fault problem across annealing time 20$\mu$s, 50$\mu$s, and 100$\mu$s. Reported values correspond to the default setting of applying no gauge, averaging over 20 gauge realizations, and the corresponding estimates with the best gauge}
\label{table:results}
\centering
\begin{tabular}{c||c|c||c|c||c|c}
\hline
 {} &  \multicolumn{2}{c|}{20 $\mu$s} &  \multicolumn{2}{c|}{50 $\mu$s} &  \multicolumn{2}{c}{100 $\mu$s} \\
\hline
{} & $R_{.99}$ & $t_{QA}$ & $R_{.99}$ & $t_{QA}$ & $R_{.99}$ & $t_{QA}$\\
\hline
No Gauge & 13953 & 0.2791 & 7193 & 0.3597 & 10231 & 1.0231\\
Average & 9227 & 0.1845 & 5655 & 0.2827 & 5819 & 0.5819\\
Best Gauge & 3174& 0.1464 & 1980 & 0.1937 & 1547 & 0.4427\\
\hline\hline
\end{tabular}
\end{table}

\section{Conclusion}\label{sec:conclusion}

In this paper, we present the first application with the route Problem $\rightarrow$ QUBO $\rightarrow$ Direct embedding into quantum hardware, where we are able to embed problem instances with sizes comparable to those found in real-world problems. For example, in the electrical circuits used for diagnosis competitions from NASA's Advanced Diagnostics and Prognostics Testbed (ADAPT), the number of components ranges between 40-100 components~\cite{kurtoglu09first}. In the present work, we were able to embed in the D-wave architecture problem instances with over 100 electrical components, including circuit breakers and sensors (see for example, last line in Table~\ref{table:embedding_HyDE} for an embedding of a network with $n_{\text{CB}}=85$ and $n_{\text{sensor}}=64$ into $n_h = 340$ hardware qubits). Key to this result is the resource-efficient construction of the problem energy function and the efficient reduction of this high-degree polynomial energy expression into the $H_{\text{QUBO}}$ (see the Appendix section for details). 

Although the tree structure of the graph allows for a more efficient mapping and embedding of the problem, and the complexity of the competition-type problem is much higher than the one presented in our first study here, we believe that we can still increase the complexity of the networks significantly while incurring a modest overhead in the additional number of qubits required. More complex networks can be obtained, for example, by increasing the number of power sources and by increasing the degree of connectivity of the network (e.g. by adding connections inside the main frame of the CB nodes or by increasing the number of CBs connected to each single sensor). We can also consider the case where we introduce expert-knowledge-based probabilities of failure for each of the components.
 
We emphasize here that our main result is not that we are able to solve a problem which no classical algorithm can solve. As the first benchmark of our studies, we picked HyDE since this was the tool available to us and capable of solving the initial set of diagnosis problems defined here. Clearly, HyDE is designed to solve problems which are more sophisticated, so it is not optimized to solve this electrical power network problem and uses memory in a non-optimal way for these problems. Also, one has to bear in mind that even though HyDE uses a smart search of the solutions space, it is still an algorithm which in the worst case scenario will try to do an exhaustive search. We believe that the cases studied here, 21 CBs and 6 faults, corresponding to a QUBO with 46 binary variables or 81 variables as implemented in the quantum machine, are still small instances for exact solvers such as akmaxsat~\cite{Kuegel2012} and/or heuristic solvers such as simulated annealing~\cite{kirkpatrick_optimization_1983,Isakov2014}. We expect in the near future to study the performance of such algorithms in finding the solution to these types of problems. Furthermore, it would be interesting to find intrinsically hard instances (for classical and quantum algorithms) like those characterized in the planning and scheduling community~\cite{Rieffel2014}. 

To the best of our knowledge there are no studies about intrinsically hard instances, such as parametrized families. The system health management community has been mostly driven by hard instances coming from practical applications, where state-of-the-art algorithms are challenged to provide the best diagnosis of real-world applications. Although we will be making developments to study those cases as well, increasing the complexity of the simple network problem described above seems also to be a natural research direction, as discussed above. 

One of the main claims in this work is the possibility of studying real complex networks fault diagnosis in future and more powerful generations of the quantum device, and the prospect of obtaining a significant speedup in cases where the assessment of the right diagnosis becomes intractable with state-of-the-art algorithms. 
In diagnosis for space applications, several subsystems can be represented as networks and hence a general network diagnosis solution can be applied to them. Some examples are
\begin{enumerate}
\item electrical power systems where batteries, power distribution components (relays, switches etc.) and loads form nodes in the node of the network and the wires connecting the components form the paths between nodes.
\item fuel loading/propulsion subsystems where tanks and valves form the nodes of the network and pipes form the paths between nodes.
\item communication and message passing subsystems where computers and memory form the nodes of the network and the communication buses form the paths between the nodes.
\end{enumerate}
In each of the subsystems a variety of sensors are typically used to measure variables at different points in the network, and these represent potential extensions to the initial work presented here.

We also showed that applying gauges to the energy functions resulting from these real-world applications, like the one discussed here, can provide a significant advantage. This was also the case in the unstructured random problems where it was originally implemented~\cite{boixo_evidence_2014}. Given the significant enhancement in performance, an open question of great importance when programming a quantum annealing device is the selection of the best gauge. This question is addressed in a follow-up publication~\citep{APO2014_DWtuning}.

It is also important to note that our quantum annealing approach not only returns the optimal solution but also returns the solutions close to this optimal solution, i.e., solutions with a number of faults higher than the minimum number of faults needed to account for the observations. This is usually not the case for diagnostics tools such as HyDE, which only report the best solution in terms of the most probable explanation. In cases where the most probable explanation does not correspond to the actual real-world solution, the quantum approach provides the next best set of candidates to choose from, without incurring any algorithmic time overhead.

\section{Appendix: Details for $H_{\text{QUBO}}$ construction}
   
The first step in the construction of $H_{QUBO}$ is to reduce the terms $g_i$, which, due to their constituent terms $f_i$ in Eq.~\ref{eq:Hconsist}, are of high locality in $H_{\text{consist}}$, to terms that are quadratic at most. From Eq.~\ref{eq:Hconsist}, each ammeter observation with $l_i=0$ has an $H_{\text{consist}}$ term of $y_i g_i = y_i f_i$ while for ammeter observations with $l_i=1$, there is a contribution to the $H_{\text{consist}}$ of the form $ y_i g_i = y_i (1-f_i)$. We deal with these two contributions using two different approaches, which are key to the compactness of our translation into QUBO. Recall that the goal is to use the fewest number of ancilla variables in the reduction of the high-degree polynomial terms $ y_i f_i$ and $-y_i f_i$ into an expression involving only quadratic terms. This quadratic expression must match the original $H_{\text{consist}}$ by being equal to zero when $l_i$ matches $f_i$, since each $g_i$ must equal zero to minimize the energy. 

A couple of important observations are useful in collapsing both of these terms efficiently. Since the optimal solution specified by the problem is the one containing the fewest number of faults, the number of faults for each $f_i$ must be at most $1$ (exactly $0$ or $1$) This is a consequence of the tree structure since, for each path, the presence of any additional fault would serve no purpose because there would be no change on the prediction of the sensor readout  associated with that path. 

Therefore, for the case $l_i=1$ (ammeter observation being HIGH), we can collapse the term $-y_i f_i$ into a quadratic term by performing the substitution $f_i = 1-D+\Sigma x_j$, where $D$ is the depth of the tree, counting all CBs from the root node to the outermost layer of CBs, i.e. $D=|P_{i}|$. This substitution, for the case of $y_i = 1$, causes the product $y_i (1-f_i)$ to take the value of zero only when $x_j=1$ for all $j$ in $P_i$. Thus, when one or more $x_j=0$, then $y_i (1-f_i) >0$, and a penalty that is an integer multiple of $\lambda_{\text{path}}$ will be incurred. 

For the case of $l_i=0$, the $f_i$ term for this problem can be remarkably rewritten without the need of additional ancilla variables as $f_i=(-D+1+\Sigma x_j)^2$. For this case of $l_i = 0$, assuming a healthy ammeter ($y_i=1$), the energy contribution of the term $y_i f_i$ should clearly be at a minimum of zero when there is at least one fault. However, as mentioned above, because the structure of the tree dictates that any additional fault in a path is redundant, therefore having two or more faults would not represent the optimal solution and could be penalized as well. Thus, the aforementioned substitution for $f_i$ ensures the constraint that $y_i f_i$ must have a minimum energy of zero when there is a single fault and higher energy in every other case. 


The steps above reduce $f_i$ to linear terms when $l_i=1$. However, whenever $l_i=0$, the term $f_i$ is a quadratic expression, so the last step in the quadratization procedure involves collapsing the now cubic terms $y_i f_i$. For this procedure we used the standard method described in Sec. 6 of Ref.~\onlinecite{Babbush2012}, along with an optimal strategy we developed for the problem at hand and which we explain next.

For $l_i=0$, the reduction of the cubic expression $y_i f_i$ to a quadratic expression consists of substitutions of the form $x_n x_m \rightarrow a_k$ or $y_i x_n \rightarrow a_k$. In general, it is more efficient to use the $y_i x_n \rightarrow a_k$ substitutions. For example, take a quaternary tree with depth $D$, and consider for the moment that there is only one path with $l_i=0$. The expansion of the quadratic term $f_i=(\Sigma x_j+1-D)^2$ results in ${D \choose 2}$ terms with an $x_n$ multiplied by a different $x_m$. 

A naive way to collapse all cubic terms to quadratic would involve all ${D \choose {2}}=D(D-1)/2$ substitutions of the form $x_n x_m \rightarrow a_k$. However, if substitutions of the form $y_i x_n \rightarrow a_k$ are made for all the $x_n$ in the path (with the exception of the root node), just $D-1$ replacements would be needed to collapse the expression to quadratic form. For example, for $l_{16}=0$, all cubic terms $\{y_{16} x_1 x_5, y_{16} x_1 x_{21}, y_{16} x_5 x_{21}\}$ can be collapsed with just $D-1=2$ substitutions $y_{16} x_{21} \rightarrow a_1$ and $y_{16} x_5 \rightarrow a_2$, changing the cubic terms into quadratic ones $\{x_1 a_2, x_1 a_1,x_5 a_1\}$. The naive contraction would require three ancillas:  $x_1 x_5 \rightarrow a_1 , x_1 x_{21} \rightarrow a_2, x_5 x_{21} \rightarrow a_3$. 

Define the variable $d \in \{1, \cdots ,D\}$ as a pointer to the $d$ layer of the tree. Also define the function $h(i,d)$ that takes the pointer and returns the label for the CB in the $i$-th path at level $d$. For example, from Figure 1a we can see that $h(1,1)=h(2,1)= \cdots = h(16,1)=1$, $h(1,2)=h(2,2)= \cdots = h(4,2)=2$, $h(5,2)=h(6,2)= \cdots = h(8,2)=3$, $h(1,3)=6$, $h(16,3)=21$, etc. where $i=1,2, \cdots ,16$ and $d=1,2,3$ for this 21 CB, 6 sensor problem.

The best strategy begins by using $y_i x_n \rightarrow a_k$ to collapse the outermost $x_{ h(i,D)}$ (nearest the sensors). This substitution is optimal because $y_i$ and $x_{h(i,D)}$ are both unique to this equation $y_i f_i$, so it is impossible for the terms to have already been collapsed from an ancilla substitution for another branch. Here, $D-1$ of the cubic terms are collapsed. The next substitution, where $x_{h(i,D-1)}$ in $y_i x_{h(i,D-1)}  \rightarrow a_k$ represents the layer of CBs nearest the outermost layer of CBs, would collapse $D-2$ terms to quadratic; this is one fewer than $D-1$ because the term with $x_{h(i,D-1)}$ multiplied by the $x_{h(i,D)}$ has already been collapsed in the previous substitution by using $y_i x_{h(i,D)} \rightarrow a_k$ so that $y_i x_{h(i,D)} x_{h(i,D-1)} \rightarrow a_k x_{h(i,D-1)}$.

Yet, making these $D-1$ substitutions using $y_i x_n \rightarrow a_k$ is not always most efficient. Depending on which paths contain $l_i=0$, there could be CB variables ($x_n$) that would be involved in many of the $y_i f_i$ to be reduced. Therefore the decision of reducing a certain pair of $x_n x_m$ vs. $y_i x_n$ for a particular $y_i f_i$ depends on how many times the variables appear globally in $H_{\text{consist}}$. Specifically, assuming $l_i=0$, 3-local terms where $x_n$ and $x_m$ are closer to the root node could be more efficiently collapsed by using the $x_n x_m \rightarrow a_k$ instead. Here, $x_n x_m$ could be non-unique to this particular $y_i f_i$ and therefore be present in multiple other branches (other equations $y_i f_i$). 

Suppose branch $i$  of the tree with its CB of depth $d$ is under consideration. Then for contractions $x_n x_m \rightarrow a_k$ and $y_i x_n \rightarrow a_k$ we can relabel $x_n$ as $x_{h(i,d)}$ and $x_m$ as $x_{h(i,d-p)}$ where $p$ represents the depth of any node along its path that is nearer the root node than $d$, i.e. $p=1, \cdots ,d-1$. The test to determine whether it is possible for the substitution $ x_{h(i,d)} x_{h(i,d-p)} \rightarrow a_k$ to be more optimal than $y_i x_{h(i,d)} \rightarrow a_k$ is the equation $d-1 \le 4^{(D-1-(d-1))}$ where base of the exponent, in this case $4$, would be $2$ for a binary tree, $3$ for a tertiary tree, etc. The left side of the equation is the number of terms that would be collapsed by performing $y_i x_{h(i,d)} \rightarrow a_k$. This is in accordance with the strategy described in the paragraph above. The right side of the equation is the number of times that terms consisting of both $x_{h(i,d)}$ and $ x_{h(i,d-p)}$ (multiplied by some $y_i$ variable) appear. Thus, only if the right side of the inequality is greater is there a chance of $x_{h(i,d)} x_{h(i,d-p)} \rightarrow a_k$ being more efficient; if this is the case, each ammeter of value $l_i=0$ that is connected to the deeper CB (the one with depth represented as $d$) must be counted, and if this number is still greater than $d-1$ (the left side of the inequality), then it will certainly be optimal to perform the substitution $x_{h(i,d)} x_{h(i,d-p)} \rightarrow a_k$ for this 3-local term. In all other cases when this test is not passed, $y_i x_{h(i,d)} \rightarrow a_k$ should be performed. 

Finally, the Hamiltonian must have terms that ensure that, for each substitution using an ancilla variable, that ancilla variable corresponds to the values of its constituent terms. For example, if the constituent terms were $x_i=1$ and $x_j=0$, then the ancilla variable must take the value of $0$. This constraint is enforced by using the methods presented Sec. V of Ref.~\onlinecite{perdomo08} or Sec. 6 of Ref.~\onlinecite{Babbush2012}.

\section*{Acknowledgment}

This work was supported by the AFRL Information Directorate under grant F4HBKC4162G001.  All opinions, findings, conclusions, and recommendations expressed in this material are those of the authors and do not necessarily reflect the views of AFRL. The authors would also like to acknowledge support from the NASA Advanced Exploration Systems program and NASA Ames Research Center.


\begin{thebibliography}{10}
\expandafter\ifx\csname url\endcsname\relax
  \def\url#1{\texttt{#1}}\fi
\expandafter\ifx\csname urlprefix\endcsname\relax\def\urlprefix{URL }\fi
\providecommand{\bibinfo}[2]{#2}
\providecommand{\eprint}[2][]{\url{#2}}

\bibitem{deKleer1987}
\bibinfo{author}{de~Kleer, J.} \& \bibinfo{author}{Williams, B.~C.}
\newblock \bibinfo{title}{Diagnosing multiple faults}.
\newblock \emph{\bibinfo{journal}{Artificial Intelligence}}
  \textbf{\bibinfo{volume}{32}}, \bibinfo{pages}{97 -- 130}
  (\bibinfo{year}{1987}).

\bibitem{narasimhan2007hyde}
\bibinfo{author}{Narasimhan, S.} \& \bibinfo{author}{Brownston, L.}
\newblock \bibinfo{title}{Hyde - a general framework for stochastic and hybrid
  modelbased diagnosis}.
\newblock In \emph{\bibinfo{booktitle}{18th International Workshop on
  Principles of Diagnosis (DX 07)}}, \bibinfo{pages}{162--169}
  (\bibinfo{year}{2007}).

\bibitem{finnila_quantum_1994}
\bibinfo{author}{Finnila, A.~B.}, \bibinfo{author}{Gomez, M.~A.},
  \bibinfo{author}{Sebenik, C.}, \bibinfo{author}{Stenson, C.} \&
  \bibinfo{author}{Doll, J.~D.}
\newblock \bibinfo{title}{Quantum annealing: A new method for minimizing
  multidimensional functions}.
\newblock \emph{\bibinfo{journal}{Chem. Phys. Lett.}}
  \textbf{\bibinfo{volume}{219}}, \bibinfo{pages}{343--348}
  (\bibinfo{year}{1994}).

\bibitem{kadowaki_quantum_1998}
\bibinfo{author}{Kadowaki, T.} \& \bibinfo{author}{Nishimori, H.}
\newblock \bibinfo{title}{Quantum annealing in the transverse ising model}.
\newblock \emph{\bibinfo{journal}{Phys. Rev. E.}}
  \textbf{\bibinfo{volume}{58}}, \bibinfo{pages}{5355} (\bibinfo{year}{1998}).

\bibitem{santoro_optimization_2006}
\bibinfo{author}{Santoro, G.~E.} \& \bibinfo{author}{Tosatti, E.}
\newblock \bibinfo{title}{Optimization using quantum mechanics: quantum
  annealing through adiabatic evolution}.
\newblock \emph{\bibinfo{journal}{J. Phys. A.}} \textbf{\bibinfo{volume}{39}},
  \bibinfo{pages}{R393--R431} (\bibinfo{year}{2006}).

\bibitem{das_2008}
\bibinfo{author}{Das, A.} \& \bibinfo{author}{Chakrabarti, B.~K.}
\newblock \bibinfo{title}{Colloquium: Quantum annealing and analog quantum
  computation}.
\newblock \emph{\bibinfo{journal}{Rev. Mod. Phys.}}
  \textbf{\bibinfo{volume}{80}}, \bibinfo{pages}{1061‚Äì21}
  (\bibinfo{year}{2008}).

\bibitem{ray1989}
\bibinfo{author}{Ray, P.}, \bibinfo{author}{Chakrabarti, B.~K.} \&
  \bibinfo{author}{Chakrabarti, A.}
\newblock \bibinfo{title}{Sherrington-kirkpatrick model in a transverse field:
  Absence of replica symmetry breaking due to quantum fluctuations}.
\newblock \emph{\bibinfo{journal}{Phys. Rev. B}} \textbf{\bibinfo{volume}{39}},
  \bibinfo{pages}{11828--11832} (\bibinfo{year}{1989}).

\bibitem{amara_global_1993}
\bibinfo{author}{Amara, P.}, \bibinfo{author}{Hsu, D.} \&
  \bibinfo{author}{Straub, J.~E.}
\newblock \bibinfo{title}{Global energy minimum searches using an approximate
  solution of the imaginary time schroedinger equation}.
\newblock \emph{\bibinfo{journal}{J. Phys. Chem.}}
  \textbf{\bibinfo{volume}{97}}, \bibinfo{pages}{6715--6721}
  (\bibinfo{year}{1993}).

\bibitem{Farhi2001}
\bibinfo{author}{Farhi, E.} \emph{et~al.}
\newblock \bibinfo{title}{A quantum adiabatic evolution algorithm applied to
  random instances of an {NP-Complete} problem}.
\newblock \emph{\bibinfo{journal}{Science}} \textbf{\bibinfo{volume}{292}},
  \bibinfo{pages}{472--475} (\bibinfo{year}{2001}).

\bibitem{santoro_theory_2002}
\bibinfo{author}{Santoro, G.}, \bibinfo{author}{Marto{\u{n}}{\'{a}}k, R.},
  \bibinfo{author}{Tosatti, E.} \& \bibinfo{author}{Car, R.}
\newblock \bibinfo{title}{Theory of quantum annealing of an ising spin glass}.
\newblock \emph{\bibinfo{journal}{Science}} \textbf{\bibinfo{volume}{295}},
  \bibinfo{pages}{2427‚Äì2430} (\bibinfo{year}{2002}).

\bibitem{brooke_quantum_1999}
\bibinfo{author}{Brooke, J.}, \bibinfo{author}{Bitko, D.},
  \bibinfo{author}{Rosenbaum, T.~F.} \& \bibinfo{author}{Aeppli, G.}
\newblock \bibinfo{title}{Quantum annealing of a disordered magnet}.
\newblock \emph{\bibinfo{journal}{Science}} \textbf{\bibinfo{volume}{284}},
  \bibinfo{pages}{779--781} (\bibinfo{year}{1999}).

\bibitem{kirkpatrick_optimization_1983}
\bibinfo{author}{Kirkpatrick, S.}, \bibinfo{author}{Gelatt, C.~D.} \&
  \bibinfo{author}{Vecchi, M.~P.}
\newblock \bibinfo{title}{Optimization by simulated annealing}.
\newblock \emph{\bibinfo{journal}{Science}} \textbf{\bibinfo{volume}{220}},
  \bibinfo{pages}{671--680} (\bibinfo{year}{1983}).

\bibitem{McGeoch2013}
\bibinfo{author}{McGeoch, C.~C.} \& \bibinfo{author}{Wang, C.}
\newblock \bibinfo{title}{Experimental evaluation of an adiabiatic quantum
  system for combinatorial optimization}.
\newblock In \emph{\bibinfo{booktitle}{Proc. of the ACM International
  Conference on Computing Frontiers}}, CF '13, \bibinfo{pages}{23:1--23:11}
  (\bibinfo{publisher}{ACM}, \bibinfo{address}{New York, NY, USA},
  \bibinfo{year}{2013}).

\bibitem{boixo_evidence_2014}
\bibinfo{author}{Boixo, S.} \emph{et~al.}
\newblock \bibinfo{title}{Evidence for quantum annealing with more than one
  hundred qubits}.
\newblock \emph{\bibinfo{journal}{Nature Physics}}
  \textbf{\bibinfo{volume}{10}}, \bibinfo{pages}{218--224}
  (\bibinfo{year}{2014}).

\bibitem{Ronnow2014_quantumspeedup}
\bibinfo{author}{R{\o}nnow, T.~F.} \emph{et~al.}
\newblock \bibinfo{title}{Defining and detecting quantum speedup}.
\newblock \emph{\bibinfo{journal}{arXiv:1401.2910}}  (\bibinfo{year}{2014}).

\bibitem{katzgraberPRX2014}
\bibinfo{author}{Katzgraber, H.~G.}, \bibinfo{author}{Hamze, F.} \&
  \bibinfo{author}{Andrist, R.~S.}
\newblock \bibinfo{title}{Glassy chimeras could be blind to quantum speedup:
  Designing better benchmarks for quantum annealing machines}.
\newblock \emph{\bibinfo{journal}{Phys. Rev. X}} \textbf{\bibinfo{volume}{4}},
  \bibinfo{pages}{021008} (\bibinfo{year}{2014}).

\bibitem{Venturelli2014}
\bibinfo{author}{Venturelli, D.} \emph{et~al.}
\newblock \bibinfo{title}{Quantum optimization of fully-connected spin
  glasses}.
\newblock \emph{\bibinfo{journal}{To be submitted}}  (\bibinfo{year}{2014}).

\bibitem{GoogleQuantumAI}
\bibinfo{author}{GoogleQuantumA.I.Lab}.
\newblock \bibinfo{title}{Where do we stand on benchmarking the {D-Wave 2}?}
\newblock
  \emph{\bibinfo{journal}{https://plus.google.com/+QuantumAILab/posts/DymNo8DzAYi}}
   (\bibinfo{year}{2014}).

\bibitem{perdomo08}
\bibinfo{author}{Perdomo, A.}, \bibinfo{author}{Truncik, C.},
  \bibinfo{author}{{Tubert-Brohman}, I.}, \bibinfo{author}{Rose, G.} \&
  \bibinfo{author}{{Aspuru-Guzik}, A.}
\newblock \bibinfo{title}{Construction of model hamiltonians for adiabatic
  quantum computation and its application to finding low-energy conformations
  of lattice protein models}.
\newblock \emph{\bibinfo{journal}{Phys. Rev. A}} \textbf{\bibinfo{volume}{78}},
  \bibinfo{pages}{012320--15} (\bibinfo{year}{2008}).

\bibitem{PerdomoOrtiz2012_LPF}
\bibinfo{author}{Perdomo-Ortiz, A.}, \bibinfo{author}{Dickson, N.},
  \bibinfo{author}{Drew-Brook, M.}, \bibinfo{author}{Rose, G.} \&
  \bibinfo{author}{Aspuru-Guzik, A.}
\newblock \bibinfo{title}{Finding low-energy conformations of lattice protein
  models by quantum annealing}.
\newblock \emph{\bibinfo{journal}{Sci. Rep.}} \textbf{\bibinfo{volume}{2}},
  \bibinfo{pages}{571} (\bibinfo{year}{2012}).

\bibitem{planningquantum}
\bibinfo{author}{Rieffel, E.} \emph{et~al.}
\newblock \bibinfo{title}{A case study in programming a quantum annealer for
  hard operational planning problems}.
\newblock \emph{\bibinfo{journal}{To be submitted}}  (\bibinfo{year}{2014}).

\bibitem{ogorman2014_bayesnet}
\bibinfo{author}{O'Gorman, B.}, \bibinfo{author}{Perdomo-Ortiz, A.},
  \bibinfo{author}{Babbush, R.}, \bibinfo{author}{Aspuru-Guzik, A.} \&
  \bibinfo{author}{Smelyanskiy, V.~N.}
\newblock \bibinfo{title}{Bayesian network structure learning using quantum
  annealing}.
\newblock \emph{\bibinfo{journal}{To be submitted}}  (\bibinfo{year}{2014}).

\bibitem{Gaitan2012}
\bibinfo{author}{Gaitan, F.} \& \bibinfo{author}{Clark, L.}
\newblock \bibinfo{title}{Ramsey numbers and adiabatic quantum computing}.
\newblock \emph{\bibinfo{journal}{Phys. Rev. Lett.}}
  \textbf{\bibinfo{volume}{108}}, \bibinfo{pages}{010501}
  (\bibinfo{year}{2012}).

\bibitem{kurtoglu09first}
\bibinfo{author}{Kurtoglu, T.} \emph{et~al.}
\newblock \bibinfo{title}{First international diagnosis competition -
  {DXC'09}}.
\newblock In \emph{\bibinfo{booktitle}{Proc. 20th International Workshop on
  Principles of Diagnosis}}, DX'09, \bibinfo{pages}{383--396}
  (\bibinfo{year}{2009}).

\bibitem{farhi2000}
\bibinfo{author}{Farhi, E.}, \bibinfo{author}{Goldstone, J.},
  \bibinfo{author}{Gutmann, S.} \& \bibinfo{author}{Sipser, M.}
\newblock \bibinfo{title}{Quantum computation by adiabatic evolution}.
\newblock \emph{\bibinfo{journal}{arXiv:quant-ph/0001106}}
  (\bibinfo{year}{2000}).

\bibitem{hogg03}
\bibinfo{author}{Hogg, T.}
\newblock \bibinfo{title}{Adiabatic quantum computing for random satisfiability
  problems}.
\newblock \emph{\bibinfo{journal}{Phys. Rev. A.}}
  \textbf{\bibinfo{volume}{67}}, \bibinfo{pages}{022314}
  (\bibinfo{year}{2003}).

\bibitem{harris2010}
\bibinfo{author}{Harris, R.} \emph{et~al.}
\newblock \bibinfo{title}{Experimental investigation of an eight-qubit unit
  cell in a superconducting optimization processor}.
\newblock \emph{\bibinfo{journal}{Phys. Rev. B.}}
  \textbf{\bibinfo{volume}{82}}, \bibinfo{pages}{024511}
  (\bibinfo{year}{2010}).

\bibitem{johnson_quantum_2011}
\bibinfo{author}{Johnson, M.~W.} \emph{et~al.}
\newblock \bibinfo{title}{Quantum annealing with manufactured spins}.
\newblock \emph{\bibinfo{journal}{Nature}} \textbf{\bibinfo{volume}{473}},
  \bibinfo{pages}{194--198} (\bibinfo{year}{2011}).

\bibitem{Barahona1982}
\bibinfo{author}{Barahona, F.}
\newblock \bibinfo{title}{On the computational complexity of ising spin glass
  models}.
\newblock \emph{\bibinfo{journal}{J. Phys. A: Math. Gen.}}
  \textbf{\bibinfo{volume}{15}}, \bibinfo{pages}{3241--3253}
  (\bibinfo{year}{1982}).

\bibitem{AlbashNJP2012}
\bibinfo{author}{Albash, T.}, \bibinfo{author}{Boixo, S.},
  \bibinfo{author}{Lidar, D.~A.} \& \bibinfo{author}{Zanardi, P.}
\newblock \bibinfo{title}{Quantum adiabatic markovian master equations}.
\newblock \emph{\bibinfo{journal}{New J. Phys.}} \textbf{\bibinfo{volume}{14}},
  \bibinfo{pages}{123016} (\bibinfo{year}{2012}).

\bibitem{Cai-14}
\bibinfo{author}{Cai, J.}, \bibinfo{author}{Macready, B.} \&
  \bibinfo{author}{Roy, A.}
\newblock \bibinfo{title}{A practical heuristic for finding graph minors}.
\newblock \emph{\bibinfo{journal}{arXiv:1406.2741}}  (\bibinfo{year}{2014}).

\bibitem{Choi2008}
\bibinfo{author}{Choi, V.}
\newblock \bibinfo{title}{Minor-embedding in adiabatic quantum computation: I.
  the parameter setting problem}.
\newblock \emph{\bibinfo{journal}{arXiv:0804.4884}}  (\bibinfo{year}{2008}).

\bibitem{Choi2011}
\bibinfo{author}{Choi, V.}
\newblock \bibinfo{title}{Minor-embedding in adiabatic quantum computation:
  {II}. minor-universal graph design}.
\newblock \emph{\bibinfo{journal}{Quantum Information Processing}}
  \textbf{\bibinfo{volume}{10}}, \bibinfo{pages}{343--353}
  (\bibinfo{year}{2011}).

\bibitem{Klymko2013}
\bibinfo{author}{Klymko, C.}, \bibinfo{author}{Sullivan, B.} \&
  \bibinfo{author}{Humble, T.}
\newblock \bibinfo{title}{Adiabatic quantum programming: minor embedding with
  hard faults}.
\newblock \emph{\bibinfo{journal}{Quantum Information Processing}}
  \bibinfo{pages}{1--21} (\bibinfo{year}{2013}).

\bibitem{APO2014_DWtuning}
\bibinfo{author}{{Perdomo-Ortiz}, A.}, \bibinfo{author}{Fluegemann, J.},
  \bibinfo{author}{Smelyanskiy, V.~N.} \& \bibinfo{author}{Biswas, R.}
\newblock \bibinfo{title}{Programming and solving real-world applications on a
  quantum annealing device}.
\newblock \emph{\bibinfo{journal}{To be submitted}}  (\bibinfo{year}{2014}).

\bibitem{harrisPRB2010}
\bibinfo{author}{Harris, R.} \emph{et~al.}
\newblock \bibinfo{title}{Experimental demonstration of a robust and scalable
  flux qubit}.
\newblock \emph{\bibinfo{journal}{Phys. Rev. B.}}
  \textbf{\bibinfo{volume}{81}}, \bibinfo{pages}{134510}
  (\bibinfo{year}{2010}).

\bibitem{Kuegel2012}
\bibinfo{author}{Kuegel, A.}
\newblock \bibinfo{title}{Improved exact solver for the weighted max-sat
  problem}.
\newblock In \bibinfo{editor}{Berre, D.~L.} (ed.)
  \emph{\bibinfo{booktitle}{POS-10}}, vol.~\bibinfo{volume}{8} of
  \emph{\bibinfo{series}{EPiC Series}}, \bibinfo{pages}{15--27}
  (\bibinfo{publisher}{EasyChair}, \bibinfo{year}{2012}).

\bibitem{Isakov2014}
\bibinfo{author}{Isakov, S.~V.}, \bibinfo{author}{Zintchenko, I.~N.},
  \bibinfo{author}{Ronnow, T.~F.} \& \bibinfo{author}{Troyer, M.}
\newblock \bibinfo{title}{Optimized simulated annealing for ising spin
  glasses}.
\newblock \emph{\bibinfo{journal}{arXiv:1401.1084}}  (\bibinfo{year}{2014}).

\bibitem{Rieffel2014}
\bibinfo{author}{Rieffel, E.~G.}, \bibinfo{author}{Venturelli, D.},
  \bibinfo{author}{Hen, I.}, \bibinfo{author}{Do, M.} \&
  \bibinfo{author}{Frank, J.}
\newblock \bibinfo{title}{Phase transitions in planning problems: Design and
  analysis or parametrized families of hard planning problems}.
\newblock \emph{\bibinfo{journal}{Accepted to AAAI-14}}
  (\bibinfo{year}{2014}).

\bibitem{Babbush2012}
\bibinfo{author}{Babbush, R.}, \bibinfo{author}{Perdomo-Ortiz, A.},
  \bibinfo{author}{O'Gorman, B.}, \bibinfo{author}{Macready, W.} \&
  \bibinfo{author}{Aspuru-Guzik, A.}
\newblock \bibinfo{title}{Construction of energy functions for lattice
  heteropolymer models: A case study in constraint satisfaction programming and
  adiabatic quantum optimization}.
\newblock \emph{\bibinfo{journal}{arXiv:1211.3422}}  (\bibinfo{year}{2012}).

\end{thebibliography}

%
%

\newpage

\end{document}